\documentclass[sigconf]{acmart}

\AtBeginDocument{%
  \providecommand\BibTeX{{%
    \normalfont B\kern-0.5em{\scshape i\kern-0.25em b}\kern-0.8em\TeX}}}

\setcopyright{acmcopyright}
\copyrightyear{2018}
\acmYear{2018}
\acmDOI{XXXXXXX.XXXXXXX}
\acmConference[Conference acronym 'XX]{Make sure to enter the correct
  conference title from your rights confirmation emai}{June 03--05,
  2018}{Woodstock, NY}
\acmPrice{15.00}
\acmISBN{978-1-4503-XXXX-X/18/06}

\usepackage{amsfonts}
\usepackage{algorithm}
\usepackage{algcompatible}

\usepackage{graphicx}
\usepackage{float}
\usepackage{subfigure}
\begin{document}

\title{Model-Agnostic Decentralized Collaborative Learning for On-Device POI Recommendation}

\author{Jing Long}
\email{jing.long@uq.edu.au}
\orcid{1234-5678-9012}
\affiliation{%
 \institution{The University of Queensland}
 \city{Brisbane}
 \state{QLD}
 \postcode{4072}
 \country{Australia}}

\author{Tong Chen}
\email{tong.chen@uq.edu.au}
\affiliation{%
 \institution{The University of Queensland}
 \city{Brisbane}
 \state{QLD}
 \postcode{4072}
 \country{Australia}}

\author{Nguyen Quoc Viet Hung}
\email{henry.nguyen@griffith.edu.au}
\affiliation{%
 \institution{Griffith University}
 \city{Gold Coast}
 \state{QLD}
 \postcode{4222}
 \country{Australia}}

 \author{Guandong Xu}
\email{Guandong.Xu@uts.edu.au}
\affiliation{%
 \institution{University of Technology Sydney}
 \city{Sydney}
 \state{NSW}
 \postcode{2007}
 \country{Australia}}

\author{Kai Zheng}
\email{zhengkai@uestc.edu.cn}
\affiliation{%
 \institution{School of Computer Science and Engineering and Shenzhen Institute for Advanced Study, University of Electronic Science and Engineering of China}
 \city{Chengdu}
 \state{Sichuan}
 \postcode{610000}
 \country{China}}
 
\author{Hongzhi Yin*}
\email{h.yin1@uq.edu.au}
\affiliation{%
 \thanks{*Corresponding author}
 \institution{The University of Queensland}
 \city{Brisbane}
 \state{QLD}
 \postcode{4072}
 \country{Australia}}

\renewcommand{\shortauthors}{Jing, et al.}

\begin{abstract}
As an indispensable personalized service in Location-based Social Networks (LBSNs), the next Point-of-Interest (POI) recommendation aims to help people discover attractive and interesting places. Currently,  most POI recommenders are based on the conventional centralized paradigm that heavily relies on the cloud to train the recommendation models with large volumes of collected users' sensitive check-in data. Although a few recent works have explored on-device frameworks for resilient and privacy-preserving POI recommendations, they invariably hold the assumption of model homogeneity for parameters/gradients aggregation and collaboration. However, users' mobile devices in the real world have various hardware configurations (e.g., compute resources), leading to heterogeneous on-device models with different architectures and sizes. In light of this, We propose a novel on-device POI recommendation framework, namely Model-Agnostic Collaborative learning for on-device POI recommendation (MAC), allowing users to customize their own model structures (e.g., dimension \& number of hidden layers). To counteract the sparsity of on-device user data, we propose to pre-select neighbors for collaboration based on physical distances, category-level preferences, and social networks. To assimilate knowledge from the above-selected neighbors in an efficient and secure way, we adopt the knowledge distillation framework with mutual information maximization. Instead of sharing sensitive models/gradients, clients in MAC only share their soft decisions on a preloaded reference dataset. To filter out low-quality neighbors, we propose two sampling strategies, performance-triggered sampling and similarity-based sampling, to speed up the training process and obtain optimal recommenders. In addition, we design two novel approaches to generate more effective reference datasets while protecting users' privacy. Extensive experiments on two datasets have shown the superiority of MAC over advanced baselines.
\end{abstract}



\received{20 February 2007}
\received[revised]{12 March 2009}
\received[accepted]{5 June 2009}

\maketitle

\section{Introduction}

Recently, the rapid growth of Location-based Social Networks (e.g., Weeplace and Foursquare) has boosted the popularity of next Point-of-Interest (POI) recommendation, which facilitates diverse applications such as mobility prediction, route planning, and location-based advertising \cite{Li2018NextPR}. In personalized POI recommendation, various state-of-the-art methods based on attentive neural networks \cite{2020Geography,2021STAN,yin2017spatial} and graph networks \cite{li2021discovering,2021Graph} have recently achieved quality recommendation performance given large volumes of historical check-in data. To support large-scale recommendation services, a powerful cloud server is commonly required to host all users' data, and to perform training and inference of the recommender, making the maintenance of such services monetarily and ecologically expensive \cite{2022Decentralized}. In addition, viewing privacy as a priority, users are increasingly reluctant to share their personal check-in trajectories with the service provider, impeding recommendation quality. Besides, in this paradigm, users' devices only act as a terminal for transmitting user data and displaying server-generated recommendations. As a result, could-based POI recommendation is overly reliant on the server's capacity and internet connectivity \cite{2020Next,yin2016adapting}, weakening its resilience.

As such, recent research resorts to on-device POI recommendations \cite{2022Decentralized,2020Next,2021PREFER} to counter the shortfalls of centralized paradigms. The core objective of this new paradigm is to deploy an accurate yet relatively lightweight recommendation model on the user side, such that recommendations can be generated in a fully on-device fashion with locally acquired user data. To fulfill this goal, a straightforward solution in the literature is based on model compression \cite{2013Estimating,2015Compressing,2015Distilling}.
In the context of on-device POI recommendation, model compression methods aim to condense a sophisticated POI recommender into a compact model to support on-device inference. For instance, a teacher-student distillation framework is utilized in \cite{2020Next} to transfer knowledge from the powerful teacher model to the compact student model. The well-trained student model is then deployed on mobile devices for all users. However, the framework is still resource-intensive due to the server's full engagement in training both the teacher and student models. Furthermore, all users are assigned the same model and there are no update mechanisms to account for the dynamics of spatial activities and diversity of user interests, leading to suboptimal performance.

Consequently, recent research is more inclined towards collaborative learning (CL) frameworks, which allow models/gradients sharing. Thus, CL is less demanding on complex model designs and more suited to the dynamic nature of POI recommendation tasks. In CL, federated learning-based POI recommenders (e.g. \cite{2021PREFER}) are a highly representative solution, where users train their models locally while retaining all private data on-device. To avoid sparsity of local data, a central server is to iteratively collect and aggregate these trained local models, and then redistribute the aggregated model to all users. However, aggregating all local models into a single global model helps the recommender to generalize, but amplifies bias towards active users' preference \cite{2021Fast}. So, further remedies are proposed in federated POI recommenders \cite{2020xw,2021rao,imran2023refrs}, which group similar users and perform group-wise aggregation to allow for more personalization of learned on-device models.

Despite the improved performance and flexibility over model compression approaches, CL-based methods still bear deficiencies for on-device POI recommendations in the real world. Firstly, a central server is always involved throughout the learning process of federated POI recommenders. This entanglement only gets stronger with the need for repetitively identifying user groups via model parameter comparison or clustering on the server. Given that, the semi-decentralized learning paradigm proposed by \cite{2022Decentralized} can be viewed as one step above federated recommenders that can significantly lower the dependency on a central server. In short, the server only needs to provide pretrained model parameters and assign similar users to the same group, which is a one-off engagement in the early stage of training. Afterward, users' on-device models are optimized by alternating local training and inter-device communication within the same user group. Unfortunately, as the intra-group collaborations require exchanging raw model parameters between users \cite{2022Decentralized}, it breaks the purposes of privacy and communication efficiency in a decentralized POI recommender. Secondly, there has been a strong assumption that user-specific knowledge can be aggregated via model parameters or gradients. However, the model homophily assumption significantly harms the practicality of such decentralized POI recommendation paradigms, as in practice, each on-device model should have a customized structure to meet the specific device capacity \cite{2021Learning}. With an exponentially increasing diversity of user devices capable of delivering POI recommendations, this assumption has to be relaxed to allow a host of structurally heterogeneous on-device models to be jointly optimized. As such, there still lacks a capable CL-based POI recommendation paradigm that can accommodate heterogeneous on-device recommenders, minimize dependency on the central server, and ensure all on-device recommenders' expressiveness without incurring privacy breaches or a heavy communication overhead.

To this end, we propose a novel on-device POI recommendation framework, namely Model-Agnostic Collaborative learning for on-device POI recommendation (MAC). In MAC, users are allowed to tailor the model configurations to suit their device capacities. As the POI embeddings are the major source of memory consumption in the recommender \cite{2020Next,2021Lightweight}, we propose to  store  embeddings only for POIs with high relevance to the current location context on the user device. The core solution to model-agnostic CL in MAC is a novel knowledge distillation (KD) scheme that facilitates knowledge exchange between heterogeneous recommendation models. Instead of sharing sensitive model parameters/gradients for equidimensional aggregations in homogeneous settings, models in MAC only share their soft decisions on a reference dataset shared among all users. In the KD process, we let models mutually incorporate knowledge from their neighbors with mutual information maximization w.r.t. the soft decisions exchanged. On the one hand, using soft decisions on the reference data bypasses the equidimensional constraint for learning decentralized models, thus scaling CL-based POI recommendation to a larger user base with heterogeneous devices and on-device models. On the other hand, the soft decisions generated from the desensitized reference data are immune to breaching user-sensitive information, which also brings a substantially small communication footprint compared with model aggregation \cite{2022Decentralized} approaches. Apparently, for each user, letting her model collaborate with all other models is computationally prohibitive and prone to lack of personalization. Given this, we propose identifying neighbors for communication regarding physical distances, category preferences, and social network distance. However, simple preliminary screening fails to filter out worthless clients and we further propose two neighbor sampling strategies, namely performance-triggered sampling and similarity-based sampling, to speed up the training process and obtain optimal recommenders. Another concern of the normal knowledge distillation is that previous works fail to consider the source of the reference datasets, and they just split the whole dataset to get the reference dataset \cite{2020Distributed,2022YE}. However, simply treating users' historical data as reference datasets has privacy concerns, and it will fail in the cold start scenario. Considering this, we design two novel approaches, namely transformative generating and probabilistic generating, to build more practical and effective reference datasets while protecting users' privacy. The primary contributions of this study are as follows:
\begin{itemize}
\item We propose a novel on-device POI recommendation framework, namely Model-Agnostic Collaborative learning for on-device POI recommendation (MAC), where users are allowed to tailor the model configurations and store embeddings only for POIs with high relevance to the current context. The core solution to model-agnostic collaborative learning (CL) is a novel knowledge distillation (KD) scheme that facilitates knowledge exchange between heterogeneous recommendation models.

\item To filter out low-quality neighbors, we design two sampling strategies, namely performance-triggered and similarity-based sampling, to speed up the training process and obtain optimal user-specific recommenders. In addition, we propose two novel approaches, namely transformative generating and probabilistic generating, which can build more practical reference datasets while protecting users' privacy.

\item We evaluate MAC with two real-world datasets. The experimental results show the effectiveness and efficiency of the proposed model in terms of privacy protection, recommendation accuracy, and the utilization efficiency of on-device computing resources.
\end{itemize}

\begin{figure*}
\vspace{-1.5em}
	\includegraphics[width=0.65\linewidth]{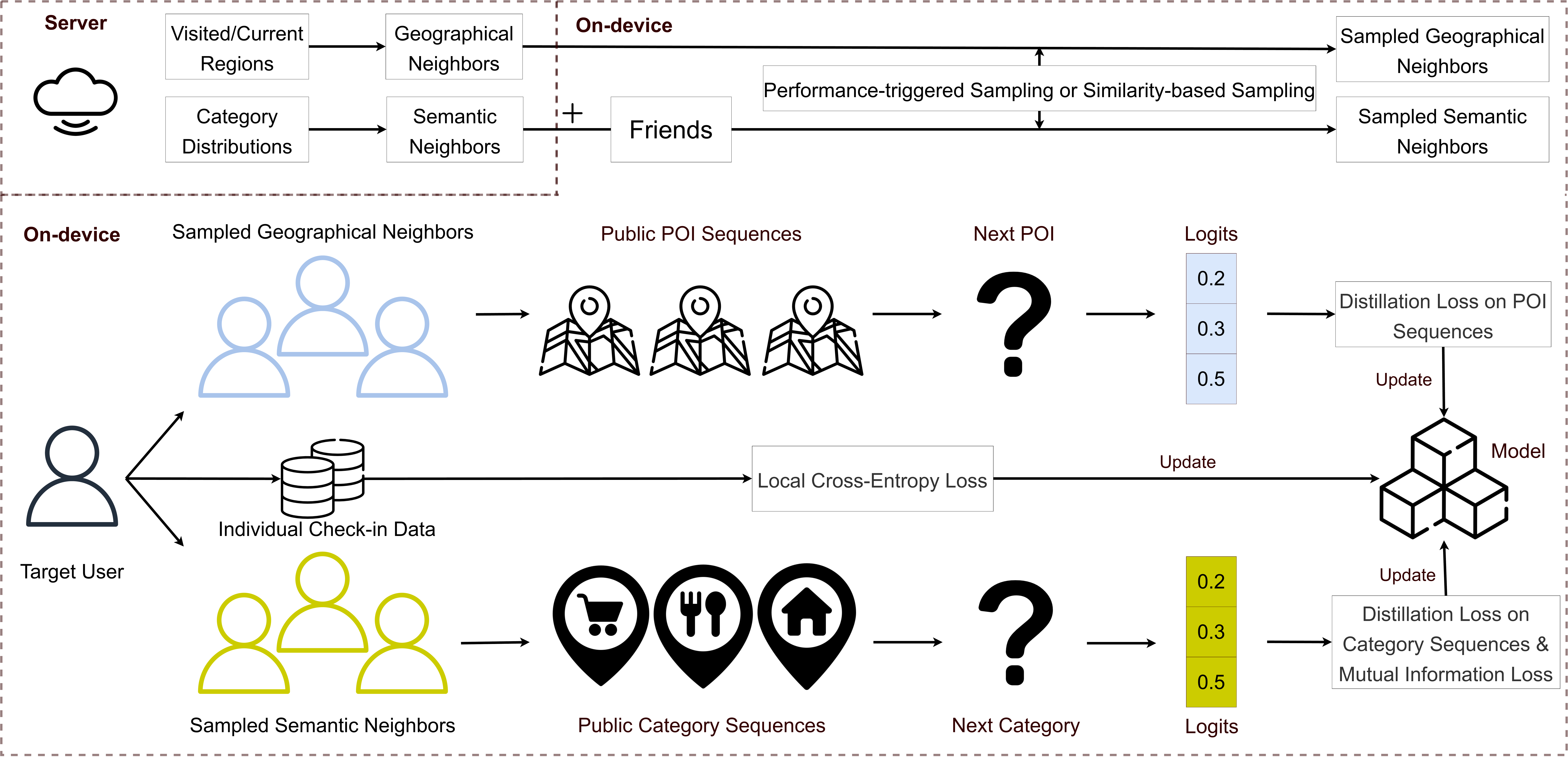}
        \vspace{-1.0em}
	\caption{The overview of our proposed MAC.}
	\label{overview} 
\vspace{-1.0em}
\end{figure*}

\section{PRELIMINARIES}\label{sec:prelim}
In this section, we first introduce important notations used in this paper and then formulate our core task.

We denote the sets of users $u$, POIs $p$ and categories $c$ as
$\mathcal{U}$, 
$\mathcal{P}$, 
$\mathcal{C}$, 
respectively. Each POI $p\in \mathcal{P}$ is associated with a category tag (e.g., entertainment or restaurant) $c_p\in \mathcal{C}$ and coordinates $(lon_p,lat_p)$.

\textbf{Definition 1: Check-in Sequence}. A check-in activity of a user indicates a user $u\in \mathcal{U}$ has visited POI $p\in \mathcal{P}$ at timestamp $t$. By sorting a user's check-ins chronologically, a check-in sequence contains $M_i$ consecutive POIs visited by a user $u_i$, denoted by $\mathcal{X}(u_i)=\{p_1, p_2,...,p_{M_i}\}$.

\textbf{Definition 2: Category Sequence}. A category sequence substitutes all POIs in the check-in sequence $\mathcal{X}(u_i)$ with their associated category tags, denoted by $\mathcal{X}^c(u_i)=\{c_{p_1},c_{p_2},...,c_{p_{M_i}}\}$.

\textbf{Definition 3: Region}. A region $r$ is essentially a geographical segment that can provide additional information about the POIs within it. Without any assumptions on predefined city districts/suburbs, we obtain a set of regions $\mathcal{R}$ by applying $k$-means clustering \cite{1967Some} on all POIs' coordinates in our paper. 

\textbf{Definition 4: Geographical Reference Dataset}. A geographical reference dataset $\mathcal{D}^{g}(r) = \{\mathcal{X}_v\}_{v=1}^{V_r}$ for region $r$ contains $V_r$ anonymous check-in sequences. Each region has its dedicated, unique reference dataset, of which the check-in activities only cover POIs in region $r$. Each user $u$ only holds region-specific reference datasets of regions she is currently or previously at.

\textbf{Definition 5: Semantic Reference Dataset}. A semantic reference dataset $\mathcal{D}^{s} = \{\mathcal{X}^c_z\}_{z=1}^Z$ contains $Z$ anonymous categorical sequences covering all POI categories $\mathcal{C}$. The semantic reference dataset is universal and shared among all users. We will detail our strategies for generating both $\mathcal{D}^{s}$ and $\mathcal{D}^{g}$ in Section \ref{sec:tech}.

\textbf{Task 1: Decentralized Next POI Recommendation}. In MAC, the roles of devices/users and the central server are defined below:
\begin{itemize}
\item \textbf{User/Device}: Each user $u_i$ holds her data $\mathcal{X}(u_i)$, $\mathcal{X}^c(u_i)$ and a structurally personalized model $\phi_i(\cdot)$ that jointly trained with the local data and enhanced by collaborating with other users. To save storage, we assume that the model $\phi_i(\cdot)$ only stores the embeddings of POIs from the regions she is currently or previously at, denoted by $r\in \mathcal{R}(u_i)$.
\item \textbf{Server}: The server is responsible for identifying neighbor sets for all users with the low-sensitivity data collected, as well as gathering reference datasets and transmitting this information back to users.

\end{itemize}
Then, by enabling model-agnostic collaborations between users, we aim to learn a performant local model for each user, which estimates a ranked list of possible POIs for her next movement.

\section{The Framework}\label{sec:tech}

This section introduces our design of the MAC framework, where an overview is depicted in Figure \ref{overview}. The main components include: 
(1) A \textbf{local objective function} that guides the optimization of each user's on-device model.
(2) A \textbf{neighbor identification process} for securely identifying cohesive user groups, operated under the cooperation between local devices and the central server.
(3) Two \textbf{data generation methods} for constructing quality yet anonymous reference datasets.
(4) A \textbf{knowledge distillation-based collaborative learning scheme} with \textbf{\textbf{neighbor sampling}} that allows refined knowledge sharing within model-agnostic user groups.

\subsection{Local Objective Function}
The main objective of MAC is to enhance personalized POI recommenders that are locally trained with users' own check-in histories. To obtain locally trained recommenders at the first place, we define the following local objective function:
\begin{equation}
  L_{loc}(u_i) = l\left(\phi_i\left(\mathcal{X}(u_i)\right),\mathcal{Y}(u_i)\right),
\end{equation}
\noindent{where} $\phi_i\left(\mathcal{X}(u_i)\right)$ is the prediction made by the recommender $\phi_i(\cdot)$ given $\mathcal{X}(u_i)$. In the POI recommendation setting, the predictions are made successively on historical POI sequences $\{p_1\}, \{p_1, p_2\},..., \{p_1,$ $p_2,...,p_{M_i-1}\}$, and $\mathcal{Y}(u_i)=\{p_2, p_3, ..., p_{M_i}\}$ is the set of corresponding ground truth POIs. $l$ is the loss function (i.e., cross-entropy in our case) to quantify the prediction error. It is worth noting that, MAC is compatible with most of the deep neural network POI recommenders as its base model, where the key innovation of our work lies in the distributed collaborative learning (CL) paradigm that coordinates with individual POI recommenders deployed on-device.

\subsection{Neighbor Identification}

To alleviate data scarcity compared with learning individual on-device recommenders in silos and reduce possible noise and overhead compared with communicating with all user devices, each user device $u_i$ in MAC is allowed to exchange knowledge with its neighbors, which are a subset of users having high affinity to $u_i$. For a robust CL process, we present two parallel strategies to identify quality neighbors for $u_i$ below.

\textbf{Geographical Neighbors}. Given the location-sensitive nature of POI recommendation tasks, users have high geographical affinity if they frequently visit venues in the same region $r\in \mathcal{R}$, thus making geographically similar users' information mutually beneficial for future movement prediction. We term such users as geographical neighbors of $u_i$, denoted by $\mathcal{G}(u_i)$. As mentioned above, with a moderate granularity of $\mathcal{R}$, we can identify a relatively dense neighbor set without risking the exposure of user privacy. Formally, we use $\mathcal{R}(u_i)=\{r^i_0,r^i_1,r^i_2,...,r^i_{R_i}\}$ to represent all regions visited by $u_i$, where $r^i_0$ denotes the user's current region (that corresponds to $p_{M_i}$) while all others are regions the user has been to. In our definition, for two users $u_i$ and $u_j$, only if $u_j$ has visited $u_i$'s current region, i.e., $r^i_0 \in \mathcal{R}(u_j)$, $u_j$ is regarded as the geographical neighbor of $u_i$. Note that, this relationship is directional, and we only identify $u_i$'s neighbors via $r^i_0$ instead of all historical regions, because the prediction on $u_i$'s next POI is mainly based on her most recent check-in $p_{M_i}$. 

\textbf{Semantic Neighbors}. Meanwhile, users' relevance is manifested in not only being physically close to each other, but also being similar at the semantic level. That is, even if being geographically distant, users can be highly relevant if their visited POIs fall into the same categories \cite{li2021discovering}, and we term such users as semantic neighbors of $u_i$, denoted by $\mathcal{S}(u_i)$. On this basis, we leverage category-level user preferences to quantify two users' semantic similarity. Formally, we use $CP(u_i)=\{\mathbb{P}(c_1),\mathbb{P}(c_2),...\mathbb{P}(c_{|\mathcal{C}|})\}$ to denote 
each user's distribution over all $|\mathcal{C}|$ POI categories, which can be easily observed from $\mathcal{X}^c(u_i)$. Then, we adopt Kullback-Leibler (KL) divergence \cite{2014Electric} to quantify the distance between two users' categorical preferences:
\begin{equation}
  d_{cat}(u_i,u_j) = KL\left(CP(u_i)\left|\right|CP(u_j)\right).
\end{equation}
For user $u_i$, we regard $h$ users with the smallest distances $d_{cat}(u_i,u_j)$ as semantic neighbors. we also account for the explicit social connections by adding users who are actual friends with $u_i$ into $\mathcal{S}(u_i)$.


Notably, in our decentralized setting, with their own check-in data, users cannot get the information to decide their geographical/semantic neighbors. As such, a central server is maintained to collect related regions $\mathcal{R}(u_i)$ and categorical preferences $CP(u_i)$ computed from $\mathcal{X}^c(u_i)$, and thus, the server can inform all users of their neighbor IDs. Compared with transmitting raw trajectories or models, the privacy risk of uploading related regions and categorical preferences is substantially lower \cite{2022Decentralized}. 

\subsection{Generating Anonymous Reference Data}\label{sec:RD}

Given the geographical neighbors $\mathcal{G}(u_i)$ and semantic neighbors $\mathcal{S}(u_i)$, the new challenge is how to extract their knowledge to enhance each user's personalized model. Although model aggregation has been proven effective in previous works \cite{2021PREFER,2022Decentralized}, it requires all participants to hold homogeneous models, and the recommendation utility is restricted by the device with the least computational resources. Thus, parameter sharing is impractical with the increasing diversity of personal devices and their varying computational capacity. As a solution, we propose a novel knowledge distillation-based CL method. Rather than transmitting gradients or weights, clients only need to share their soft decisions on desensitized reference datasets with their neighbors. Essentially, with the same reference dataset, those generated soft decisions are a compact characterization of each on-device model. Besides, compared with exchanging gradients/models, the use of soft decisions can substantially lower the communication cost and is far less sensitive. 

Previous CL frameworks \cite{2020Distributed,2022YE} have tried assigning the same reference dataset across all users to facilitate such model-agnostic knowledge exchange, but they are subject to strong limitations in POI recommendation tasks. Firstly, the reference dataset is supposed to cover all regions' POIs, allowing all users to produce probability distributions that are mutually comparable. Apparently, this will harm the quality of soft decisions, since each user's local model is learned with geographically constrained data, and can hardly make reliable predictions on POIs outside familiar regions. Secondly, storing and processing check-in sequences covering a full set of POIs is a heavy burden for user devices, marking down the practicality of MAC. Thirdly, despite the important role of the reference datasets in CL, a common practice to obtain them \cite{2020Distributed,2022YE} is to draw a subset of real samples from the user data, which negatively affects user privacy and thus hurts real-world usability. 

\textbf{Geographical and Semantic Reference Datasets.} For knowledge sharing among geographical neighbors, to guarantee each local model's prediction quality over the reference dataset, we prepare one reference dataset for each region, where each region-specific reference dataset $\mathcal{D}^{g}(r) = \{\mathcal{X}_v\}_{v=1}^V$ contains $V$ anonymous check-in sequences in region $r$. Given this, user $u_i$ can communicate with her geographical neighbors by comparing their soft decisions w.r.t. each check-in sequence in $\mathcal{D}^{g}(r^i_0)$ (recall that $r^i_0$ is $u_i$'s current region). Also, as now each $\mathcal{D}^{g}(r)$ only contains region-specific POIs, the amount of POI information needed to be stored is substantially reduced. Unfortunately, this strategy is inapplicable to CL with semantic neighbors, as $u_i$ and her semantic neighbors might be geographically distant from each other, and their locally trained models may exhibit divergent behaviors on the same region-specific reference dataset. Hence, instead of predicting the actual POIs, we let them predict high-level categorical transitions on a semantic reference dataset $\mathcal{D}^{s}= \{\mathcal{X}^c_z\}_{z=1}^Z$ with $Z$ anonymous categorical sequences, defined in Section \ref{sec:prelim}. Analogously, knowledge sharing between $u_i$ and her semantic neighbors is enabled by computing soft decisions upon $\mathcal{D}^{s}$. Given that only categorical transitions are modeled with this dataset and the number of POI categories is relatively small, sharing one universal semantic reference dataset will meet both quality and memory efficiency demands. 

Meanwhile, the challenge from privacy aspect still needs further attention. To maintain all users' privacy when building reference datasets, we resort to pseudo data as a natural solution. Specifically, we put forward two data generation strategies, namely transformative generation and probabilistic generation, which allow MAC to produce sensible yet anonymous trajectories. We introduce both strategies below, which are adopted alternatively in MAC. 

\textbf{Transformative Generation}. Instead of directly publishing users' check-in sequences, we propose to obfuscate them by a series of transformations, which are widely used for augmenting time series data \cite{2016Data,2018Data,2021Data}. In a nutshell, we mix and match two users' check-in/category sequences via a decompose-recompose pipeline, where the revamped sequences differ from but imitate original ones. As the raw check-ins are exchanged in this process, we only allow this to happen between two users $u_i$ and $u_m$ if (1) they are mutual friends; and (2) they have visited at least one same POI, which can be judged once $u_m$ sends her check-in sequence to $u_i$. Specifically, with two check-in sequences $\mathcal{X}(u_i)$, $\mathcal{X}(u_j)$, we decompose each of them into two subsequences by slicing from a common POI $p^{ij}\in \mathcal{X}(u_i)\cap \mathcal{X}(u_j)$, i.e., $\mathcal{X}'(u_i) = \{p^i_1, ..., p^{ij}\}$ and $\mathcal{X}''(u_i) =\{p^{ij}, ..., p^i_{M_i}\}$ for $\mathcal{X}(u_i)$, and $\mathcal{X}'(u_j) = \{p^j_1, ..., p^{ij}\}$ and $\mathcal{X}''(u_j) = \{p^{ij}, ..., p^j_{M_j}\}$ for $\mathcal{X}(u_j)$. Then, we join $\mathcal{X}'(u_i)$ with $\mathcal{X}''(u_j)$, and $\mathcal{X}'(u_j)$ with $\mathcal{X}''(u_i)$ to form two new sequences $\mathcal{X}^{g}_1$ and $\mathcal{X}^{g}_2$, where a repetitive $p^{ij}$ is deleted from both sequences. For both newly generated sequences, we specify their region(s) $r$ as the most frequently visited one, and discard POIs out of region $r$ to guarantee all generated POI sequences are region-specific. Analogously, the universal semantic reference dataset $\mathcal{D}^{s}$ can be constructed via transformative generation on users' category sequences, while the additional step of region-specific processing is not required. 


\textbf{Probabilistic Generation}. Though only perturbed data is used in transformative generation, it originates from users' raw historical data. To meet stricter privacy restrictions and deal with regions with limited check-ins, we propose an alternative that only leverages the statistics from users' insensitive category sequences for data generation, which we call probabilistic generation. Firstly, we calculate the conditional probabilities of all categories based on the category sequences submitted by all users. Formally, we use $P(c_n)=\{\mathbb{P}(c_n|c_1),\mathbb{P}(c_n|c_2),...,\mathbb{P}(c_n|c_m),...,\mathbb{P}(c_n|c_{|C|})\}$ to represent all conditional probabilities for $c_n$, and each $\mathbb{P}(c_n|c_m)$ is calculated as:
\begin{equation}
  \mathbb{P}(c_n|c_m)=\frac{\mu(c_n|c_m)}{\sum_{m'=1}^{|C|}\mu(c_n|c_{m'})},
\end{equation}
\noindent{where} $\mu(c_n|c_m)$ denotes the number of occurrences of $c_m$ right after $c_n$. Then, a pseudo category sequence $\mathcal{X}^{s}$ can be generated by: (1) selecting a category randomly as the starting point; and then (2) repeatedly deciding subsequent categories according to the conditional probabilities of its previous category. 
By iteratively generating multiple $\mathcal{X}^{c}$, we can create a semantic reference dataset $\mathcal{D}^{s}$ that cover all categories. The next step is to generate POI sequences based on the semantic reference dataset. Within each region, we randomly sample a sequence of POIs that correspond to the category sequence $\mathcal{X}^{c}\in \mathcal{D}^{s}$, where we further restrict the distance between any two consecutive POIs to be less than 5km. 

\subsection{Collaboration via Knowledge Distillation}
With the reference dataset ready, we design a novel knowledge distillation-based CL protocol to extract knowledge from soft decisions received from geographical neighbors $\mathcal{G}(u_i)$ and semantic neighbors $\mathcal{S}(u_i)$.

\textbf{Collaborative Learning with Geographical Neighbors}. Given user $u_i$ and the geographical reference dataset of her current region $\mathcal{D}^{g}(r^i_0)$, the CL with her geographical neighbors $\mathcal{G}(u_i)$ is facilitated by minimizing their disagreement of soft decisions on $\mathcal{D}^{g}(r^i_0)$, which is quantified as: 
\begin{equation}\label{eq:Lgeo}
  L_{geo} = \frac{1}{|\mathcal{G}(u_i)|}\sum_{u_j\in \mathcal{G}(u_i)}\bigg{(} \sum_{\mathcal{X} \in \mathcal{D}^{g}(r^i_0)}\left|\left|\phi_i
  \left(\mathcal{X}\right)-\phi_j(\mathcal{X})\right|\right|^2_2 \bigg{)},
\end{equation}

\noindent{where} $\phi_i(\cdot)$ and $\phi_j(\cdot)$ respectively denote the local recommenders possessed by $u_i$ and her neighbors.

\textbf{Collaborative Learning with Semantic Neighbors}. Similarly, we can also align the soft decisions on the semantic reference dataset $\mathcal{D}^{s}$ between $u_i$ and $\mathcal{S}(u_i)$. However, as only category tags are recorded in the reference dataset and hence predictions can only be made on categorical preferences, we cannot directly update the POI embeddings in $\phi_i(\cdot)$ by comparing users' soft decisions. 
Instead, we design a two-step strategy to acquire knowledge from semantic neighbors. Firstly, we update the category embeddings in $\phi_i(\cdot)$ via the following: 
\begin{equation}\label{eq:Lcat}
  L_{cat} = \frac{1}{|\mathcal{S}(u_i)|}\sum_{u_j\in \mathcal{S}(u_i)}\bigg{(}\sum\limits_{\mathcal{X}^c \in \mathcal{D}^{s}}\left|\left|\psi_i
  \left(\mathcal{X}^c\right)-\psi_j(\mathcal{X}^c)\right|\right|^2_2 \bigg{)},
\end{equation}

\noindent{where} $\psi_i(\cdot)$ is a lightweight predictor that infers a probability distribution over all categories from a historical category trajectory $X^c\in \mathcal{D}^{s}$.
Hereby, since the embeddings of a POI $p$ and its associated category $c_p$ can be treated as two views of the POI, we further update POI embeddings by maximizing the mutual information (MI) between these two views:
\begin{equation}
  L_{MI} = -\sum_{\forall p} \log\frac{\exp({f(p, c_{p}))}}{\sum_{\forall c'\neq c_p} 
  \exp(f(p,c'))},
\end{equation}

\noindent{where} $f(\cdot)$ is a function of the pairwise similarity between the POI and the category. In MAC it is implemented by a bilinear network that takes the POI and category embeddings as its input:
\begin{equation}
  f(p,c)=\sigma(\textbf{e}_p^{\top}\textbf{W}\textbf{e}_{c}),
\end{equation}
\noindent{where} $\textbf{e}_p\in\mathbb{R}^{d}$ and $\textbf{e}_c\in\mathbb{R}^{d}$ are POI and category embeddings, and $\textbf{W}\in\mathbb{R}^{d\times d}$ is a learnable weight matrix. By optimizing the loss function, we can maximize the mutual information between positive pairs while minimizing that between negative pairs. Then, the final loss for collaborative learning with semantic neighbors is defined as the ensemble of both parts:
\begin{equation}
  L_{sem} = L_{cat} + L_{MI}.
\end{equation}

\renewcommand{\algorithmiccomment}[1]{\textbf{#1}}
\newcommand{\algorithmicnewcomment}[1]{\textit{#1}}
\begin{algorithm}[t]
  \caption{Optimizing MAC. All processes are implemented on user's side unless specified.}
  \label{alg:A}
\begin{algorithmic}[1]
    \STATEx /*Server-side engagement starts*/
        \STATE Server receives $R(u_i)$, $CP(u_i)$ from all $u_i\in U$;
        \STATE Generate $\mathcal{G}(u_i)$ and $\mathcal{S}(u_i)$ w.r.t. $R(u_i)$ and $CP(u_i)$ for all $u_i\in \mathcal{U}$;
        \STATE Compose $\mathcal{D}^{g}(r)$ for all $r\in \mathcal{R}$ and $\mathcal{D}^{sem}$ via transformative or probabilistic generation; 
        \STATE Send $\mathcal{G}(u_i)$, $\mathcal{S}(u_i)$, $\mathcal{D}^{g}(r^i_0)$, $\mathcal{D}^{s}$ to each $u_i \in U$;
    \STATEx /*Server engagement ends*/
    \FOR[in parallel]{$u_i\in\mathcal{U}$}
        \STATE Impute $\mathcal{S}(u_i)$ with $u_i$'s social friends;
        \STATE Initialize $\phi_i(\cdot)$, $\mathcal{G}'(u_i)$, $\mathcal{S}'(u_i)$ for all $u_i\in\mathcal{U}$;
        \REPEAT
               \STATE Receive soft decisions on $\mathcal{D}^{g}(r^i_0)$, $\mathcal{D}^{s}$ respectively 
               \STATEx \hspace{0.55cm} from $\mathcal{G}'(u_i)$, $\mathcal{S}'(u_i)$;
               \STATE Receive soft decisions on $\mathcal{D}^c$ from $N^{s}_{cat}(u_i)$;
                \STATE Take a gradient step w.r.t. $L_{loc}+\gamma(\mu L_{geo}+(1-\mu)L_{sem})$ 
                \STATEx \hspace{0.55cm} to update $\phi_i(\cdot)$;
                \STATE Update $\mathcal{G}'(u_i)$, $\mathcal{S}'(u_i)$ with performance-triggered or 
                \STATEx \hspace{0.55cm} similarity-based sampling;
         \UNTIL{convergence}
    \ENDFOR

\end{algorithmic}
\end{algorithm}
\vspace{-1.5em}

\subsection{Learning Local Recommenders with Dynamic Neighbor Sampling}
The optimization workflow of MAC is presented in Algorithm \ref{alg:A}. At the very beginning, we perform neighbor identification and prepare all reference datasets on the server side (lines 1-3). Then, the server sends the corresponding reference datasets and neighbor information back to all users (line 4), where its engagement halts. On the device side (lines 5-14), each user will iteratively update its local recommender based on the synergic loss $L_{loc}+\gamma(\mu L_{geo}+(1-\mu)L_{sem})$, where $\gamma$ and $\mu$ jointly controls the strength of different knowledge distillation components.

Meanwhile, while the neighbor identification process provides $u_i$ with abundant users that are likely to contribute to learning $\phi_{i}(\cdot)$, not all users in $\mathcal{G}(u_i)$/$\mathcal{S}(u_i)$ are necessarily valuable and might conversely dilute the local model's quality. On top of the influence of noise, it is also inefficient to communicate with all predefined neighbors in every optimization step. Hence, we further propose two neighbor sampling strategies to draw a small subset of $\alpha$ users $\mathcal{G}'(u_i) \in \mathcal{G}(u_i)$ and $\mathcal{S}'(u_i) \in \mathcal{S}(u_i)$ for CL. By swapping the full neighbor sets $\mathcal{G}(u_i)$ and $\mathcal{S}(u_i)$ in Eq.(\ref{eq:Lgeo}) and Eq.(\ref{eq:Lcat}) to $\mathcal{G}'(u_i)$ and $\mathcal{S}'(u_i)$, we are able to speed up the training process and obtain optimal recommenders. 

\textbf{Performance-triggered Sampling}. 
Intuitively, picking neighbors that provide high-quality knowledge is beneficial, where the fluctuations in the local loss function $L_{loc}(u_i)$ is a strong indicator on whether or not the knowledge distilled from $u_i$'s neighbors is sufficiently informative. So, 
in every CL epoch $o$, if the change in $\Delta L_{loc}^o(u_i)$ falls below a predefined threshold $\tau$, the previously sampled neighbors $\mathcal{G}'(u_i)$/$\mathcal{S}'(u_i)$ will be redrawn from $\mathcal{G}(u_i)$/$\alpha$. In our work, we define $\Delta L_{loc}(u_i)$ in the following ratio format: 
\begin{equation}
  \Delta L_{loc}^o(u_i)=\frac{|L^{o}_{loc}(u_i) - L^{o-1}_{loc}(u_i)|}{L^{o-1}_{loc}(u_i)} \times 100\%,
\end{equation}
\noindent{where} we can adjust $\tau$ to actively control the sensitivity to performance change, and we empirically adopt $\tau=1\%$ for both neighbor types in our work. Note that we draw samples uniformly at random with replacements to avoid repetition.

\textbf{Similarity-based Sampling}. Naturally, another feasible solution is to narrow down the neighbors to communicate in $\mathcal{G}(u_i)$/$\mathcal{S}(u_i)$ on the go by similarity measures. To achieve this, we utilize the soft decisions w.r.t. the reference datasets by comparing them in every CL iteration via KL divergence \cite{2014Electric}:
\begin{equation}
  d_{soft}(u_i,u_j) = \sum_{\forall X \in \mathcal{D}^{ref}}KL
  \left(\phi_i \left(X\right)\left|\right|\phi_j\left(X\right)\right),
\end{equation}
\noindent{where} $\mathcal{D}^{ref} = \mathcal{D}^{g}(r^i_0)$ if $u_j\in \mathcal{G}'(u_i)$, and $\mathcal{D}^{ref} = \mathcal{D}^{s}$ if $u_j\in \mathcal{S}'(u_i)$. Then, for user $u_i$, $\beta$ users with the smallest $d_{soft}(u_i,u_j)$ will be respectively selected from $\mathcal{G}(u_i)$ and $\mathcal{S}(u_i)$. 

In MAC, we employ either similarity-based or performance-triggered sampling as they substitute each other. Also, as all models are changing over time, the sampling process dynamically adjusts $\mathcal{G}'(u_i)$ and $\mathcal{S}'(u_i)$, as shown in line 12 of Algorithm \ref{alg:A}.

\section{Experiments}

In this section, we conduct extensive experiments on two real-world datasets to evaluate the effectiveness and efficiency of MAC.
    

\begin{table}[htbp]
\vspace{-1.0em}
  \centering
  \caption{Dataset statistics.}
  \label{table:A}
   \vspace{-1.0em}
    \begin{tabular}{lrr}
          \hline
          & \multicolumn{1}{r}{Foursquare} & \multicolumn{1}{r}{Weeplace} \\
    \hline
    \#users & 7,507     & 4,560 \\
    \hline
    \#POIs & 80,962     & 44,194 \\
    \hline
    \#categories & 436     & 625 \\
    \hline
    \#check-ins & 1,214,631     & 923,600 \\
    \hline
    \#check-ins per user & 161.80   & 202.54 \\
    \hline
    \end{tabular}%
    \vspace{-1.0em}
\end{table}%

\begin{table*}[t]
  \centering
  \caption{Recommendation performance comparison with baselines.}
  \label{table:B}
  \vspace{-1.0em}
    \begin{tabular}{|l|cccc|cccc|}
      \hline
          & \multicolumn{4}{c|}{Foursquare}  & \multicolumn{4}{c|}{Weeplace} \\
          \cline{2-9}
          & \multicolumn{1}{c}{HR@5} & \multicolumn{1}{c}{NDCG@5} & \multicolumn{1}{c}{HR@10} & \multicolumn{1}{c|}{NDCG@10} & \multicolumn{1}{c}{HR@5} & \multicolumn{1}{c}{NDCG@5} & \multicolumn{1}{c}{HR@10} & \multicolumn{1}{c|}{NDCG@10} \\
          \hline
          MF    & 0.0835 & 0.0598 &	0.0962 & 0.0669	& 0.1045 & 0.0612	& 0.1293 & 0.0872 \\
          LSTM  & 0.1920 & 0.1227 &	0.2812 & 0.1648 &	0.2215 & 0.1356 &	0.3301 & 0.1578 \\
          STAN  & 0.2914 & 0.1756 &	0.4157 & 0.2524 &	0.3205 & 0.1846 &	0.4503 & 0.2783 \\
          LLRec & 0.2771 & 0.1488 &	0.3406 & 0.1908 &	0.2912 & 0.1781 &	0.3613 & 0.2255 \\
          PREFER & 0.2895 & 0.1696 &	0.3607 & 0.2183 &	0.3055 & 0.1874 &	0.3707 & 0.2359 \\
          DCLR  & 0.3078 & 0.1797 &	0.4297 & 0.2653 &	0.3304 & 0.1912 &	0.4664 & 0.2838 \\
          D-Dist  & 0.2257 & 0.1193 &	0.2915 & 0.1621 &	0.2474 & 0.1343 &	0.3581 & 0.1933 \\
          SQMD  & 0.2818 & 0.1471 &	0.4054 & 0.2357 &	0.2953 & 0.1531 &	0.4342 & 0.2360\\
          MAC-P & \textbf{0.3138} & \textbf{0.1865} & \textbf{0.4319} & \textbf{0.2712} & \textbf{0.3414} & \textbf{0.2032} & \textbf{0.4843} & \textbf{0.2937} \\
          MAC-S & 0.3055 & 0.1843 &	0.4257 & 0.2637 &	0.3365 & 0.1988 &	0.4734 & 0.2845\\
          \hline
        \end{tabular}%
  \label{tab:addlabel}%
\end{table*}%

\subsection{Datasets and Evaluation Protocols}

We adopt two real-world datasets to evaluate our proposed MAC, namely Foursquare \cite{2020Will} and Weeplace \cite{2013Personalized}. Both datasets include users' check-in histories in the cities of New York, Los Angeles, and Chicago. Following \cite{Li2018NextPR,2018Content}, users and POIs with less than 10 interactions are removed. Table \ref{table:A} summarizes the statistics of the two datasets. Among those datasets, 10\% of the check-in sequences are randomly selected as the reference datasets with the constraint of covering all POIs. The two strategies to generate reference datasets are also processed based on those selected check-in activities. Then, inspired by \cite{2018Neural,2019Enhancing}, we employ the leave-one-out protocol for evaluation. Specifically, for each of the remaining check-in sequences, the last check-in POI is for testing, the second last POI is for validation, and all others are for training. In addition, the maximum sequence length is set to 200. For each ground truth, instead of ranking all e-commerce products \cite{2020On}, we only pair it with 200 unvisited and nearest POIs within the same region of the sequence as the candidates for ranking. The rationale is, different from e-commerce products \cite{2020On}, in the scenario of POI recommendations that are location-sensitive, users seldom travel between two POIs consecutively that are far away from each other \cite{2022Decentralized,li2021discovering}. On this basis, we leverage two ranking metrics, namely Hit Ratio at Rank $k$ (HR@$k$) and Normalized Discounted Cumulative Gain at Rank $k$ (NDCG@$k$) \cite{2007CoFiRank} where HR@$k$ only measures the times that the ground truth is present on the top-$k$ list, while NDCG@$k$ cares whether the ground truth can be ranked as highly as possible.

\subsection{Baselines and Experimental Setting}

We compare MAC with both the centralized and on-device POI recommenders: 

\begin{itemize}
\vspace{-0.2em}
  \item \textbf{MF} \cite{2014GeoMF}: It is a classic centralized POI recommendation system based on user-item matrix factorization. 
  
  \item \textbf{LSTM} \cite{1997Long}: This recurrent neural network can capture short-term and long-term dependencies in sequential data.
  
  \item \textbf{STAN} \cite{2021STAN}: It learns explicit spatiotemporal correlations of check-in trajectories via a bi-attention approach.
  
  \item \textbf{LLRec} \cite{2020Next}: It utilizes the teacher-student training strategy to obtain the compressed model that can be deployed locally.
  
  \item \textbf{PREFER} \cite{2021PREFER}: This federated POI recommendation paradigm allows the server to collect and aggregate locally trained models, as well as redistribute the federated model.
  
  \item \textbf{DCLR} \cite{2022Decentralized}: This decentralized collaborative learning framework allows locally trained models to share knowledge between homogeneous neighbors by model aggregation.
  
  \item \textbf{D-Dist} \cite{2020Distributed}: It aims at letting locally trained models with random heterogeneous neighbors via comparing their soft decisions on a public reference dataset.
  
  \item \textbf{SQMD} \cite{2022YE}: It is also a decentralized distillation framework where neighbors are defined by comparing their responses on the shared reference dataset.
\vspace{-1.0em}
\end{itemize} 

\subsection{Experimental Settings}

As mentioned above, this work is compatible with almost any centralized POI recommendation model. To achieve advanced accuracy, we exploit STAN \cite{2021STAN} as the base model. Besides, since this work supports heterogeneous structures, we randomly assign the latent dimension $d\in\{8,16,32,64,128\}$ to users and each one makes up 20\%. This setting is also applied to other heterogeneous frameworks (i.e., D-Dist and SQMD). For fairness, all other baselines are evaluated with the above dimensions separately and the final results are averaged. Additionally, in this work, each city is divided into 5 regions by applying k-means clustering which is discussed in Section \ref{sec:prelim}. It is worth noting that in MAC, we implement both the performance-triggered and similarity-based approaches for neighbor sampling, which are respectively marked as MAC-P and MAC-S.

For hyperparameters, we set $\alpha$ to $5$, $\beta$ to $10$, $\gamma$ to $0.5$ and $\mu$ to $0.7$. The impacts of the above four hyperparameters will be further discussed in Section \ref{sec:hyperparam}. Apart from this, we set $h$ to 50, and adopt a learning rate of $0.002$, dropout of $0.2$ on all deep layers, batch size of $16$, and the maximum training epoch is set to $50$.

\begin{table*}
    \centering
    \begin{minipage}{\linewidth}
        \begin{minipage}{0.7\linewidth}
            \centering
            \makeatletter\def\@captype{table}\makeatother\caption{Model size (kb) and accuracy among decentralized POI recommenders.}
            \label{table:C}
            \setlength\tabcolsep{3pt}
            \vspace{-1.0em}
            \begin{tabular}{|l|cc|cc|cc|cc|cc|}
                \hline
                  & \multicolumn{2}{c|}{d = 8} & \multicolumn{2}{c|}{d = 16} & \multicolumn{2}{c|}{d = 32} & \multicolumn{2}{c|}{d = 64} & \multicolumn{2}{c|}{d = 128} \\
                \cline{2-11}
                  & \multicolumn{1}{c}{Size} & \multicolumn{1}{c|}{HR@10} & \multicolumn{1}{c}{Size} & \multicolumn{1}{c|}{HR@10} & \multicolumn{1}{c}{Size} & \multicolumn{1}{c|}{HR@10} & \multicolumn{1}{c}{Size} & \multicolumn{1}{c|}{HR@10} & \multicolumn{1}{c}{Size} & \multicolumn{1}{c|}{HR@10} \\
                \hline
                DCLR  & 1130  & 0.3674 & 2252  & 0.4413 & 4502  & 0.5417 & 9020  & 0.5165 & 18127 & 0.4651 \\
                D-Dist & 1057  & 0.2863 & 1963  & 0.3309 & 3852  & 0.3514 & 7968  & 0.3925 & 15564 & 0.3293 \\
                SQMD  & 1157  & 0.3637 & 2054  & 0.4194 & 3648  & 0.4845 & 8245  & 0.4585 & 16362 & 0.4449 \\
                MAC-P & 232   & 0.4057 & 462   & 0.4678 & 694   & 0.5404 & 1316  & 0.5114 & 3186  & 0.4963 \\
                MAC-S & 226   & 0.4026 & 479   & 0.4341 & 742   & 0.5416 & 1276  & 0.5010 & 2987  & 0.4877 \\
                \hline
            \end{tabular}%
        \end{minipage}
        \begin{minipage}{0.3\linewidth}
        \centering
        \makeatletter\def\@captype{table}\makeatother\caption{Traning time (in hours).}
        \label{table:D}
        \setlength\tabcolsep{3pt}
        \renewcommand\arraystretch{1.165}
        \vspace{-1.0em}
            \begin{tabular}{|l|c|c|}
                \hline
                      & \multicolumn{1}{c|}{Foursquare} & \multicolumn{1}{c|}{Weeplace} \\
                    \hline
                    DCLR  & 76 & 57 \\
                    D-Dist & 24 & 19 \\
                    SQMD  & 51 & 39 \\
                    MAC-P & 40 & 32 \\
                    MAC-S & 29 & 21 \\
                \hline    
            \end{tabular}%
        \end{minipage}
    \end{minipage}
\end{table*}

\subsection{Recommendation Effectiveness}

The performance comparison among all the POI recommenders is summarized in Table \ref{table:B}, where we observe the following findings.

Among the centralized POI recommenders, LSTM is superior to MF on both datasets due to the effective use of short-term and long-term dependencies of sequential check-in activities. Besides, thanks to spatiotemporal correlations of consecutive and non-consecutive check-in activities, STAN has higher accuracy than LSTM. Compared with the advanced baseline STAN, our method still yields highly competitive results. This is because the centralized STAN model is trained with check-ins across multiple cities where knowledge learned from one city might be noisy for the recommendation tasks of other cities, leading to inferior performance of STAN. Instead, our work can achieve better personalization, and learn more expressive models with the collaborative learning architecture. 

In the meantime, MAC outperforms all on-device POI recommenders on both datasets in terms of all metrics. Specifically, LLRec has the worst performance since all personal data is not included in the training process, ignoring individual preferences, while our method can trade negligible privacy risks for large improvement in accuracy. Although PREFER has a noticeable improvement in accuracy, it still needs to collect and aggregate users' personalized models, exposing user privacy to potential breaches. Thus, MAC is more capable of providing both more accurate recommendations and stronger privacy protection with less reliance on the cloud. 

\subsection{Efficiency Analysis}
Amid all homogeneous on-device recommenders, DCLR achieves the best performance due to its effective collaborative learning strategies. However, it suffers from the constraint that all participants must hold homogeneous models for model aggregation. In contrast, MAC can not only support users having varied dimensions like D-Dist and SQMD, but also allows user devices effectively utilize limited storage resources by storing partial POI embeddings. To prove the above view, with respect to all heterogeneous on-device recommenders, we record the averaged model size (kb) and recommendation accuracy (HR@10 on Weeplace) for the latent dimensions $d\in\{8,16,32,64,128\}$. The results are shown in Table \ref{table:C}, where we can observe that the average model size of MAC is far less than that of the other three on-device recommenders among all dimensions. Additionally, MAC still outperforms DCLR and other heterogeneous models in terms of recommendation accuracy at all dimensions, showing that MAC can make more efficient use of on-device resources to provide more performant recommendations. 

Besides, our proposed strategies for neighbor identification and neighbor sampling can speed up the training process. We record the training time till model convergence, where the results are shown in Table \ref{table:D}. We can observe that D-Dist has the least training time since each device in the network will communicate with a static group of peers at each iteration. In exchange, D-Dist has the worst recommendation accuracy. Beyond that, MAC converges faster than all other models while providing more accurate recommendations. Interestingly, compared to MAC-P, MAC-S achieves higher recommendation accuracy, but it needs more training time as it tends to change the neighbor set composition more frequently. Instead, MAC-S only communicates with a relatively stable neighbor set, thus converging faster but being prone to local optima.

\subsection{Analysis on Reference Data Generation}

To validate the efficacy of two proposed strategies for generating reference datasets described in Section \ref{sec:RD}, we evaluate the recommendation accuracy of both MAC-P and MAC-S on original, transformative generated, and probabilistic generated reference datasets. The results are shown in table \ref{table:E}. Apparently, the performance on the original reference datasets is similar to the transformative generated reference datasets. This proves the efficacy of the transformative strategy considering the strong privacy protection. Meanwhile, although the performance on the probabilistic reference datasets 
lightly drops, it is still capable for the region without enough historical check-in activities. Please note all other experiments are based on transformative generated reference datasets.

\begin{table}[t]
  \centering
  \caption{Effect of reference dataset sources on recommendation accuracy (HR@10 is demonstrated).}
  \vspace{-1.0em}
  \label{table:E}
    \begin{tabular}{|l|cc|cc|}
    \hline
          & \multicolumn{2}{c|}{Foursquare} & \multicolumn{2}{c|}{Weeplace} \\
    \cline{2-5}
          & \multicolumn{1}{c}{MAC-P} & \multicolumn{1}{c|}{MAC-S} & \multicolumn{1}{c}{MAC-P} & \multicolumn{1}{c|}{MAC-S} \\
    \hline
    Original & 0.4357 & 0.4232 & 0.4901 & 0.4698 \\
    Transformative & 0.4319 & 0.4257 & 0.4843 & 0.4734 \\
    Probabilistic & 0.4173 & 0.4129 & 0.4715 & 0.4581 \\
    \hline
    \end{tabular}%
  \label{tab:addlabel}%
\end{table}%

\subsection{Hyperparameter Sensitivity}\label{sec:hyperparam}

\begin{figure*}
        \includegraphics[width=0.70\linewidth]{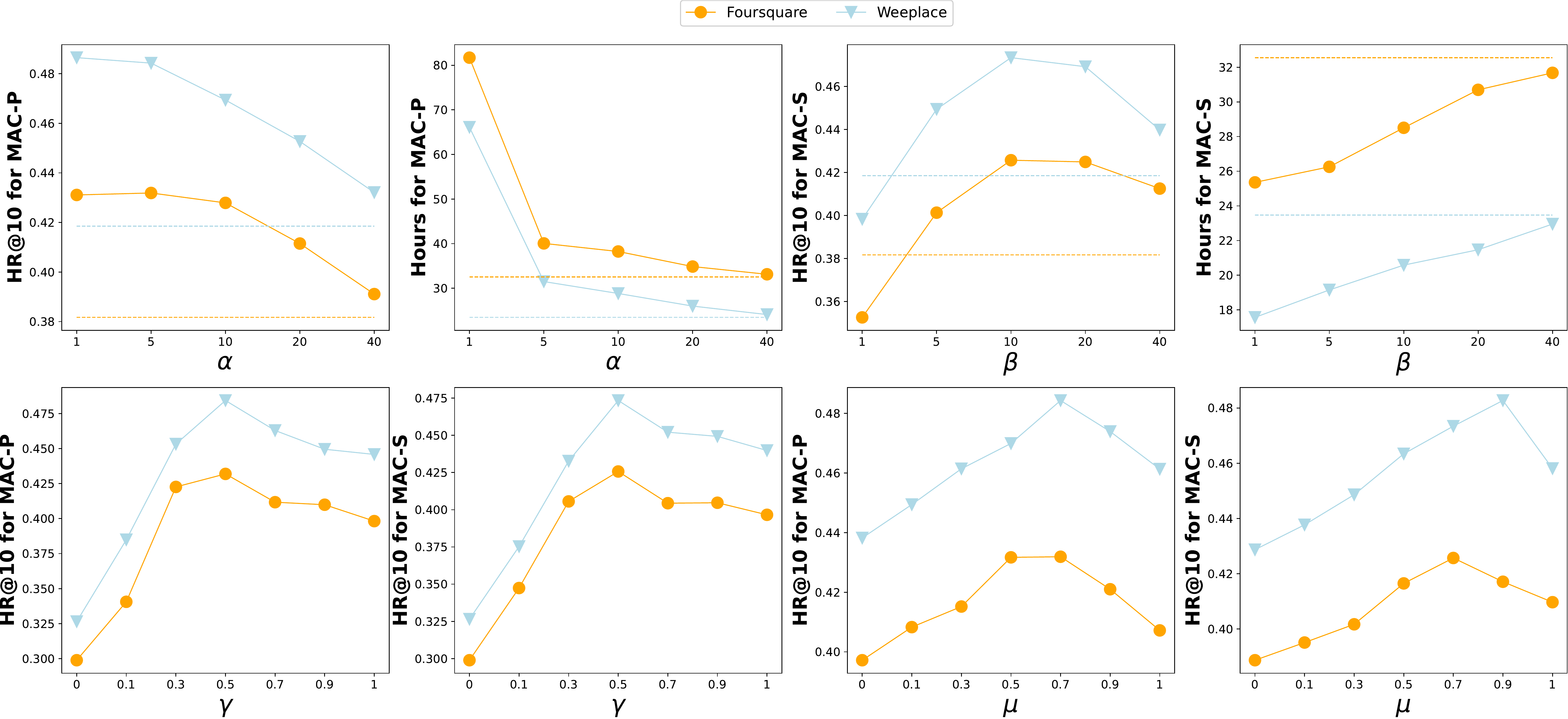}
    \vspace{-0.4cm}
	\caption{Hyperparameter sensitivity. The horizontal lines indicate MAC variants without any sampling strategy.}
	\label{F} 
\end{figure*}

In this section, we illustrate the effect of four hyperparameters on the performance of MAC.
Specifically, we evaluate the sample size in performance-triggered sampling $\alpha$ on the accuracy and training time of MAC-P,
the sample size in similarity-based sampling $\beta$ on the accuracy and training time of MAC-S,
the weight $\gamma$ that controls the knowledge from neighbors on the accuracy of MAC-P and MAC-S,
and the trade-off $\mu$ between geographical and semantic neighbors for communication on the accuracy of MAC-P and MAC-S.
The results are shown in Figure \ref{F}.

\textbf{Impact of $\alpha$.} We experiment on a series of $\alpha\in\{1, 5, 10, 20, 40\}$. In general, a larger $\alpha$ harms the performance of MAC-P on both datasets, as low-quality neighbors are included, but it requires less training time with lower communication frequencies. In the extreme case of communicating with all predefined neighbors (dotted line), we record the lowest accuracy, which proves the effectiveness of performance-triggered sampling. 

\textbf{Impact of $\beta$.} We study the impact of $\beta\in\{1, 5, 10, 20, 40\}$. As $\beta$ increases from $1$ to $10$, there is a generally upward trend in MAC-S's accuracy with more valuable knowledge. However, it starts to decrease when $\beta$ exceeds $10$ as low-quality neighbors are included. Meanwhile, the training time keeps rising since more neighbors are processed. Compared with the extreme case (dotted line), MAC-S achieves higher accuracy with less training time, showing the effectiveness of similarity-based sampling.

\textbf{Impact of $\gamma$.} $\gamma$ is evaluated in $\{0,0.1,0.3,0.5,0.7,0.9,1\}$. Apparently, the lowest accuracy is obtained if users fail to share knowledge with neighbors ($\gamma=0$), showing the significance of the collaborative learning framework. However, the performance will decline if knowledge from neighbors has an excessive proportion ($\gamma>0.5$).

\textbf{Impact of $\mu$.} $\mu$ is examined in $\{0,0.1,0.3,0.5,0.7,0.9,1\}$. The best performance is observed when $\mu=0.7$ in most cases, showing more importance of geographical factors compared to semantic factors.

\section{Related Work}

This section reviews recent literature on related areas including centralized models for POI recommendation, on-device frameworks for POI recommendation, and distributed knowledge distillation.

\subsection{Next POI Recommendation}

To help people discover attractive places by analyzing user-POI interactions, early models mainly focused on matrix factorization \cite{2014GeoMF} and Markov chains \cite{2013Where,zheng2016keyword}. Recently, models based on recurrent neural networks (RNN) have been proven effective in capturing spatiotemporal dependencies among POI sequences \cite{yin2013lcars, chen2020sequence, 2018DeepMove, yin2016spatio, chen2020try}. Meanwhile, SGRec \cite{li2021discovering} constructs graph-augmented POI sequences to fully capture the collaborative signals from semantically correlated POIs and mine sequential properties, which achieves better accuracy than RNN-based models. In addition, attentive neural networks \cite{2021STAN, 10.1145/3477495.3531983,chen2019air,yin2015joint} utilize self-attention layers to capture relative spatiotemporal information of all check-in activities along the sequence. It is worth noting that all the above models are cloud-based, leading to undeniable problems of privacy issues and high demand for powerful cloud. Instead, MAC lies in the distributed learning paradigm to provide secure and stable services.
\subsection{On-device POI Recommendation}

On-device frameworks have been proven effective in addressing most shortcomings of cloud-based learning for POI recommendations. Intuitively, Wang et al. \cite{2020Next} deployed compressed models on mobile devices for secure and stable POI recommendations. To maintain the robustness of the whole on-device framework, local compressed models inherit the knowledge from the teacher model which is trained with public data. Then, Guo et al. \cite{2021PREFER} proposed a federated learning framework for POI recommendations, allowing edge servers to collect and aggregate locally trained models, and send the aggregated model back to all users. Apparently, the above models are still highly dependent on the cloud server. On this basis, Jing et al. \cite{2022Decentralized} proposed a semi-decentralized learning paradigm with device collaboration that allows user devices to learn and combine knowledge from two types of neighbors. However, all aforementioned collaborative learning-based POI recommenders hold a strong assumption that all on-device models must share an identical design, allowing user-specific knowledge sharing via parameters/gradients aggregation. In contrast, MAC supports heterogeneity in model structures.

\subsection{Distributed Knowledge Distillation}

Knowledge distillation was initially proposed in model compression \cite{2019Towards}, aimed at extracting knowledge from a powerful teacher model trained with mass data, to improve the performance of the light student model. The information is transferred by minimizing the disagreement between teacher and student models on an unlabeled reference dataset. However, such a way of knowledge transmission is not restricted to the teacher-student pattern, where \cite{2020Online} and \cite{2021Adversarial} have proven the effectiveness of exploiting knowledge distillation to jointly train student models without any teacher model. This is also applicable in knowledge sharing between heterogeneous device networks where user devices are allowed to have different model architectures \cite{2020Distributed, 2022YE}. Our work is the first one to exploit knowledge distillation for knowledge sharing between heterogeneous POI recommendation models.

\section{Conclusion}

To support heterogeneous architecture in decentralized POI recommendations where users are allowed to have a customized structure to meet the specific device capacity, we incorporate mutual information maximization into knowledge distillation to exchange information between similarity-based groups decided by physical distances, category distributions, and social networks. To remove low-quality neighbors, during the training process, we design two strategies to further sample neighbors according to their performance or similarity to the target user, reducing training time and obtaining optimal recommenders. Apart from this, we propose two novel approaches to build more practical reference datasets while providing strong privacy protection. Experimental results have rigorously demonstrated the efficacy of MAC, revealing its ability to provide excellent services in decentralized POI recommendations.

\begin{acks}
This work is supported by the Australian Research Council under the streams of Future Fellowship (Grant No. FT210100624), Discovery Project (Grant No. DP190101985), Discovery Early Career Research Award (Grant No. DE200101465 and No. DE230101033), and Industrial Transformation Training Centre (No. IC200100022). It is also supported by NSFC (No. 61972069, 61836007 and 61832017), Shenzhen Municipal Science and Technology R\&D Funding Basic Research Program (ICYJ 20210324133607021), and Municipal Government of Quzhou under Grant No. 2022D037.
\end{acks}
\bibliographystyle{ACM-Reference-Format}
\bibliography{sample-sigconf}

\end{document}